\documentclass[aps,twocolumn,nofootinbib,groupedaddress,showpacs,epsf]{revtex4}
\usepackage{graphicx}
\usepackage{psfrag}
\usepackage{color}

\def\a{\alpha}

\def\d{\delta}
\def\e{\epsilon}
\def\ve{\varepsilon}
\def\g{\gamma}
\def\r{{\bf r}}

\def\k{\kappa}

\def\l{\lambda}

\def\s{\sigma}
\def\t{\theta}
\def\vp{\varphi}

\def\S{{\bf S}}

\begin{document}
\bibliographystyle{apsrev}


\title{Geometry-dependent electrostatics near 
contact lines}



\author{Tom Chou}
\affiliation{Dept. of Biomathematics, UCLA, Los Angeles, CA 90095-1766}


\date{\today}

\begin{abstract}
Long-ranged electrostatic 
interactions in electrolytes modify their
contact angles on charged substrates in a scale and geometry dependent manner.  
For angles measured at scales smaller than the typical Debye screening
length, the wetting geometry near the contact line must be explicitly considered.  Using
variational and asymptotic methods, we derive new transcendental equations for the
contact angle that depend on the electrostatic potential only at the three phase
contact line. Analytic expressions
are found in certain limits and compared with predictions for contact angles measured
with lower resolution. An estimate for electrostatic contributions to 
{\it line} tension is also given.
\end{abstract}

\pacs{47.10.+g, 68.08.-p, 68.08.Bc}

\maketitle


Modern microfluidic and patterning applications call for directed fluid flow and
wetting on treated surfaces often exhibiting  complex surface chemistry
\cite{CHA95,PET92,TROIAN,NAPL}.  Mechanisms of differential wetting of small droplets
of electrolytes have also been exploited as electrically activated switches and
micropumps \cite{CJKIM}.  In such systems, the effects of surface ionization have been
found to be important \cite{CHA95,NAPL,KAYSER,DIGILOV,FOKKINK}.  

Consequently, one often considers two immiscible fluids of dielectric constants
$\e_{0},\e_{1}$ that partially wets an ionizable, rigid substrate, as shown in Figs.  1. 
The two liquids, depending on their individual pKa's, will differentially
hydrolyze/ionize substrates such as glass.  Ionizable surfactants may also be adsorbed,
imparting a fixed, relatively uniform surface charge $\sigma$ at the interfaces. 
Microscopically, surface tensions $\g$ arise from mismatches in short-ranged molecular
interactions ({\it e.g.} van der Waal's) among the various species.  Electrostatic double
layers also contribute to surface energies. The application of the classical Young-Dupr\'{e}
(Y-D) equation \cite{YDREF}, $\g_{01}'\cos\alpha^{*} \approx \g_{0s}'-\g_{1s}'$, with
electrostatically modified surface energies ({\it e.g.} $\g_{1s}'= \g_{1s}+{1\over
2}\s_{1}\vp_{1}$, where $\s_{1}$ and $\vp_{1}$, are surface charges and potentials in,
say, liquid 1 far from the contact point $P$) is accurate provided the apparent contact
angle $\a^{*}$ is measured at a point $P^{*}$ outside the range $\kappa^{-1}$ of the
electric double layers.  However, the screening length $\kappa^{-1}$, although typically
smaller than the resolution of optical goniometry measurements, can be within the
resolution (nanometers) of emerging angle measurement techniques using AFM \cite{POMPE}. 
Even at lower resolutions, finite-sized double-layer effects may be relevant for contact
angle measurements. For example, hydrolysis of pure water gives $\kappa^{-1}\sim 1\mu$m,
while in organic mixtures, with fewer mobile ions, the screening length can be even
longer \cite{NAPL}. If the contact angle is measured at $P$, within the ionic double layers, 
the simple Y-D equation is not appropriate. 

\begin{figure}
\includegraphics[height=3.8in]{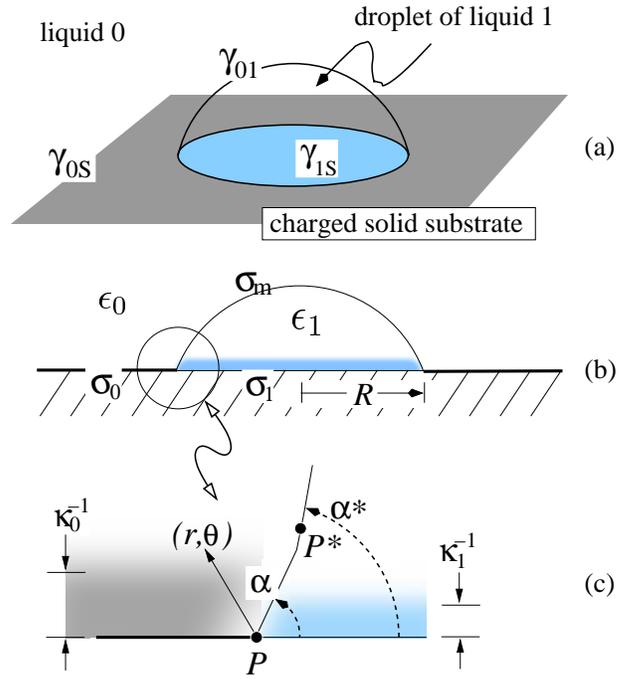}
\caption{(a) Two liquids with dielectric constants $\e_{0},\e_{1}$ wets a charged substrate. 
(b) The substrate is
ionized and acquires fixed charge densities of $\s_{0}, \s_{1}$.  Charged surfactants
can also impart a surface charge $\s_{m}$ at the fluid-fluid interface. (c)
In order to satisfy boundary conditions, the
different screening layers deform near the contact point $P$.}
\label{fig1}
\end{figure}

Previous theories that consider surface energy modifications
\cite{CHA95,KAYSER,DIGILOV,FOKKINK,BUTKUS,WHITE,JPH}, have either assumed
microscopic-ranged interactions or infinite-system surface free energy changes ({\it
i.e.} $\g \rightarrow \g + {1 \over 2}\s\vp$).  Effects of the wedge-like geometry on
the intrinsic electrostatics near the contact line have not been considered.  In this
letter, we derive formulae for the angle\footnote{Measured within both double layers,
but outside the length scale of other more microcscopic interactions.} at the true
three phase contact line in the presence of long-ranged electrostatic interactions. 
Rich features arise in this most simple and classic problem when geometry is
self-consistently incorporated. We propose using weak ionic solutions as a system with
precisely controllable electrostatics for studying charge contributions to contact
angle and line tension forces.  Our results are summarized by Eqns.  (\ref{YDC}),
(\ref{RESULT1}), and (\ref{T0}).  


The ``mechanical'' free energy of an axisymmetric liquid droplet of footprint
radius $R$ (cf. Figs. 1a,b) in contact with a flat, solid substrate is $G_{mech} =
(\g_{1s}-\g_{0s})\int d\r+\g_{01}\oint d{\bf S}_{01}+ G_{body}$, where $\g$ are
the surface tensions {\it sans} electrostatic interactions, $d\r$ is the in-plane
surface element, $\S$ is the liquid-liquid surface element. The $G_{body}$ term may
include gravitational energy, $(\rho_{1}-\rho_{0}) g/2 \int h^{2}(r)d\r$, and/or
Lagrange multipliers to {\it e.g.} fix droplet volumes of incompressible
liquids.  The electrostatic free energy for a specified surface charge ensemble,
written as a functional of the local electrostatic potential
$\vp$ is \cite{CHOU}

\begin{equation}
\begin{array}{ll}
\displaystyle G_{e\ell} = \oint\sigma(\S)\vp(\S)d\S -
\int d\r dz\left[{\e(\r,z)\over 8\pi} \vert\nabla\vp\vert^{2}+U[\vp]\right]
\label{GELL}
\end{array}
\end{equation}


\noindent where $U[\vp(\r)]\equiv \sum_{i}e\nu_{i}c_{i}^{\infty}
(\exp(-\nu_{i}\vp(\r))-1)$ is a term summing the interactions among mobile
charged species $i$, each with valency $\nu_{i}$ and bulk concentration
$c_{i}^{\infty}$.  We have expressed all energy and length quantities in terms of
$k_{B}T$ and the Bjerrum length $\ell_{B}=e^{2}/k_{B}T$.  Variation of
$G_{e\ell}[\vp(\r), \nabla\vp(\r); h(\r), \nabla h(\r)]$ with respect to
$\vp(\r)$ for a fixed droplet height function $h(\r)$ yields the
Poisson-Boltzmann equation with appropriate boundary conditions. Similarly, variation
of $G_{T}=G_{mech}+G_{e\ell}$ with respect to the droplet height $h(\r)$
determines the complete shape of the electrolyte droplet via

\begin{equation}
\gamma_{01}\partial^{2}_{{\bf t}} h_{\perp}(\r) =
\left[{\e\over 8\pi}\vert\nabla\vp\vert^{2}+U[\vp]\right]^{0}_{1},
\label{H}
\end{equation}

\noindent where $h_{\perp}$ is the deformation normal to the tangent 
${\bf t}$ relative to a constant slope. 
Further minimizing the boundary terms in $G_{T}$
(which are independent of $U[\vp]$) with respect to the 
position of the contact point,
$\delta G_{T}(z=0)/\delta R = 0$, yields\footnote{The liquid-liquid 
surface tension $\g_{01}$ may also change due to 
electrocapillarity, but this can be 
modification measured independently 
using {\it e.g.} pendant drop methods.}

\begin{equation}
\g_{01}\cos\alpha = (\g_{0s}-\g_{1s}) + (\s_{0}-\s_{1})\vp(P;\alpha).
\label{YDC}
\end{equation}

The first two terms arise from setting the variations in the boundary terms of
$G_{mech}$ to zero and reproduces the Young-Dupr\'{e} equation.  The last term in
(\ref{YDC}) is a new generalization of the Y-D equation and arises from
minimizing the boundary terms of $G_{e\ell}$. The additional term
depends only on the jump in the solid surface charge and the electrostatic
potential $\vp(P)$ at the three phase contact line $P$. This potential is
found by solving the Poisson-Boltzmann equation $\nabla^{2}\vp(\r) = U'[\vp(\r)]$
in the appropriate geometry, subject to boundary conditions.  Therefore, $\vp(P)$
will depend parametrically on the droplet shape (and hence the contact angle
$\a$) near $P$.  Equation (\ref{YDC}) is {\it exact} provided the electrostatic
energy is given by $G_{e\ell}$, and gives an implicit formula for predicting the
contact angle $\alpha$.

In the following, we compute $\vp(P;\a)$ in the linearized limit $(U[\vp] \approx
\kappa^{2}\vp^{2}/2)$, valid for $e\vp(\r)/k_{B}T \ll 1$, by solving Kelvin's ({\it aka}
Debye-H\"{u}ckel) equation in each of the two fluid domains depicted in Figs. 1: 

\begin{equation}
\displaystyle \Delta \vp_{j} = \kappa_{j}^{2}\vp_{j}(r,\t) 
\quad \mbox{in fluid}\, j=0,1.
\label{KELVIN}
\end{equation}

\noindent The screening lengths $\kappa_{j}^{-1} \equiv \left({4\pi e\over
\epsilon_{j}}\sum_{i}\nu_{i}^{2}c_{i}^{\infty}\right)^{-1/2}$ are assumed much
smaller than the dimensions of the droplet.  Furthermore, neglecting gravity,
(\ref{H}) shows that $h_{\perp}(\r)$ varies (relative to a perfect wedge) over a
length scale $L_{h}\sim \sqrt{\g_{01}/(\kappa \vp^{2}(P))}$. Provided $L_{h} \gg
\kappa^{-1}$, the wedge is distorted only in the region where $\vp \approx 0$,
sufficiently beyond the contact point $P$ to be appreciably influenced by
electrostatic interactions.  Thus, $\vp(P)$ will be computed using a perfect wedge
geometry. The boundary conditions associated with (\ref{KELVIN}) in 2D wedge
domains are

\begin{equation}
\begin{array}{c}
\displaystyle \e_{1}\partial_{\theta}\vp_{1}(r,0) = 4\pi r\s_{1} \\
\displaystyle \e_{0}\partial_{\theta}\vp_{0}(r,\pi) = -4\pi r\s_{0} \\
\displaystyle \vp_{0}(r,\a)-\vp_{1}(r,\a) =0 \\
\displaystyle \e_{0}\partial_{\theta}\vp_{0}(r,\a)- 
\e_{1}\partial_{\theta}\vp_{1}(r,\a) = 4\pi r \s_{m}.
\label{BC}
\end{array}
\end{equation}

\noindent The linear problem defined above is related to the classic problem of wave
scattering from a wedge, which remains a substantial mathematical and
computational challenge \cite{RAWLINS}. The problem is best attacked using the
Lebedev-Kantorovich (LK) integral transform 

\begin{equation}
f_{j}(x,\theta) = \int_{0}^{\infty}
\vp_{j}(r,\t)K_{ix}(\k_{j} r){dr \over r}\equiv {\cal L}_{\kappa_{j}}\vp_{j},
\label{LK}
\end{equation}

\noindent and its inverse

\begin{equation}
\vp_{j}(r,\theta) = {2\over \pi^{2}}
\int_{0}^{\infty}f(x,\t)K_{ix}(\k_{j} r)x\mbox{sh} (\pi x)dx\equiv
{\cal L}^{-1}_{\kappa_{j}}f_{j}.
\label{LKI}
\end{equation}

\noindent If the contact angle can be measured sufficiently close the contact point
(closer than $\k_{j}^{-1}$, such as in \cite{POMPE}), the potential $\vp(\k r\rightarrow P)$ can be easily
expressed in terms of its transform $f(x,\theta)$.  We will henceforth scale distance as $\xi
= \kappa_{0}r$ and reinsert $\kappa_{0}$ back into the final, quoted results. 
Asymptotic analysis of the integral representation $K_{ix}(\xi) =
(1/2)\int_{-\infty}^{\infty} \exp\left[-\xi\mbox{ch} t + ixt\right]dt$ yields
$\lim_{\xi\rightarrow 0}K_{ix}(\xi) \sim (\sin xt^{*})/x$, where $t^{*} =
\lim_{\xi\rightarrow 0} \cosh^{-1}(1/\xi) = \infty$. The limit $\lim_{t^{*}\rightarrow
\infty} (\sin xt^{*})/x = \pi\delta(x)$, implies 

\begin{equation}
\lim_{\xi \rightarrow 0}\vp(\xi,\theta) = \lim_{x\rightarrow 0}
\left[x^{2}f(x,\theta) \right]. 
\label{VP0}
\end{equation}

\noindent We find analytic expressions for $\vp(r,\theta)$ in two
limiting cases that illustrate the full range of behaviors: $\k_{0} \approx \k_{1}$,
and $\k_{1} \gg \k_{0}$. In the former case of nearly identical 
screening lengths unequal dielectric constants $\e_{0}\neq \e_{1}$
provide an nontrivial electric field jump across $\theta = \a$.  In the latter case,
we will choose a small screening length ({\it e.g.}, high salt) in region 1 without
loss of generality.  In both cases, the potential can be expanded in the power series
$\vp_{j}(r,\t) = \sum_{n=0}^{\infty}\mu^{(n)}(\ve)\vp_{j}^{(n)}(r,\t)$, where
$\ve$ is a small parameter that depends on the ratio of screening lengths and the
relevant regime: 

{\it $\k_{1}\approx \k_{0}$ limit - }In this limit, $\mu^{(n)}(\ve) = 
\ve^{n}$ and $\ve\equiv
(\k_{1}/\k_{0})^{2}-1$. At each order $\ve^{n}$ the governing equations
are $\Delta\vp_{0}^{(n)} = \vp_{0}^{(n)}$ for $\t\geq\a$, and
$\Delta\vp_{1}^{(n)} = \vp_{1}^{(n)}+\vp_{1}^{(n-1)}$ for $\t\leq\a$.  
Upon applying ${\cal L}_{1}$ to the $n^{th}$ order Debye-H\"{u}ckel equations,
$\partial_{\t}^{2}f_{0}^{(n)} = x^{2}f_{0}^{(n)}(x,\t)$ and
$\partial_{\t}^{2}f_{1}^{(n)} = x^{2}f_{1}^{(n)}(x,\t)=g^{(n-1)}(x,\t)$ where
$g^{(n-1)}(x,\t) \equiv {\cal L}_{1}\vp^{(n-1)}r^{2}(1-\d_{n,0})$.


Similarly, the transformed boundary conditions at each order become
$\partial_{\t}f_{0}^{(n)}(x,\t=\pi)=(2\pi^{2}\s_{0}/\e_{0})\mbox{sech}(\pi
x/2)\delta_{n,0}$, $\partial_{\t}f_{1}^{(n)}(x,\t=0)=-(2\pi^{2}\s_{1}/\e_{1})
\mbox{sech}(\pi x/2)\delta_{n,0}$, and $\e_{1}\partial_{\t}f_{1}^{(n)}(x,\t=\a)-
\e_{0}\partial_{\t}f_{0}^{(n)}(x,\t=\a) = 2\pi^{2}\s_{m}\mbox{sech}(\pi x/2)\delta_{n,0}$. 
Since $f_{1}^{(0)}(x,\theta)$ obeys a homogeneous equation, 
it is determined explicitly. 
The equations for $f_{1}^{(n)}(x,\t)$ can be solved with the appropriate Green function
$G(x,\t,\t') = -\mbox{ch}\, x\t_{<}\mbox{ch}\,x(\t_{>}-\a)/(x\mbox{sh}\,x\a),
\partial_{\t'}G(x,\t,\t'=0,\a)=0$, where $\t_{<}(\t'_{>})$ is the smaller(larger) of
$\t,\t'$.  We obtained, after some algebra, an integral equation for 
$f_{1}(x,\theta)$. The first iteration of this integral equation yields

\begin{equation}
\ve f_{1}^{(1)}(x,\theta) = \ve\int_{0}^{\infty}\!\!dx'\int_{0}^{\alpha}d\t'\,
H(\t,\t';x,x')f_{1}^{(0)}(x',\t')
\label{INTEGRAL}
\end{equation}



\noindent where 

\begin{equation}
\begin{array}{l}
\:\hspace{-4mm}\displaystyle H(\t,\t';x,x') \equiv \\[13pt]
\displaystyle {(\e_{1}/4)G(\t,\t';x)(x^{2}-x'^{2})x'\mbox{sh}\pi x'\mbox{sh}x\a \over
\left(\e_{1}\mbox{sh}x\a + \e_{0}\mbox{ch}x\theta\,\mbox{th}x(\pi-\a)
\right)\mbox{sh}{\pi\over 2}(x+x')
\mbox{sh}{\pi \over 2}(x-x')}.
\end{array}
\end{equation}

\noindent  Upon integrating 
(\ref{INTEGRAL}) over $\theta'$, and using $f_{1}(x,\theta) \approx
f_{1}^{(0)}(x,\theta)+\ve f_{1}^{(1)}(x,\theta)$ in (\ref{VP0}), we find $\vp(P;\a)$ up to 
$O(\ve)$:

\begin{equation}
\begin{array}{l}
\displaystyle \vp(P;\a) \sim {2\pi^{2}/\kappa_{0}\over
\e_{1}\a+\e_{0}(\pi-\a)}\bigg[(\s_{0}+\s_{1}+\s_{m}) \\[13pt]
\:\hspace{3mm} + \ve\big(\s_{0}S_{0}(\a;\l)
+\s_{1}S_{1}(\a;\l)+\s_{m}S_{m}(\a;\l)\big) + O(\ve^{2})\bigg].
\end{array}
\label{RESULT1}
\end{equation}

\begin{figure}
\includegraphics[width=3.3in]{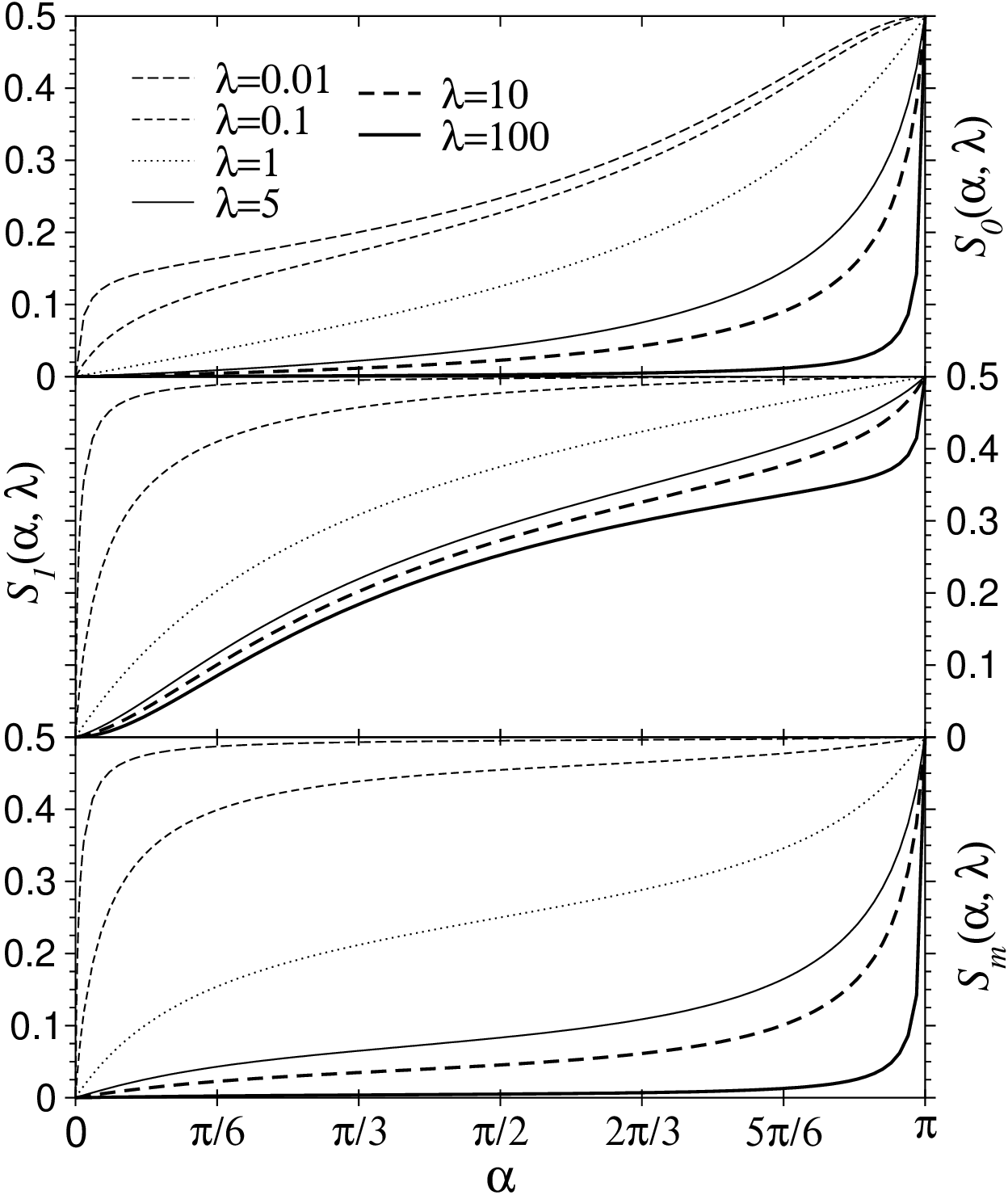}
\caption{The functions $S_{i}(\a;\l=\e_{0}/\e_{1})$ representing the order 
$\ve = \kappa_{1}^{2}/\kappa_{0}^{2}-1$ contributions to 
$\vp(\a;P)$ (cf. Equation (\ref{RESULT1})).}
\label{S}
\end{figure}

\noindent The first (zeroth order) term arises from 
$\lim_{x\rightarrow 0} x^{2}f_{1}^{(0)}$, while 
the $O(\ve)$ terms can be expressed as single $\a$-dependent integrals
$S_{i}(\a;\lambda)$. The functions $S_{i}$ are plotted {\it vs.} $\a$ in Fig. \ref{S} for 
various dielectric mismatches $\e_{0}/\e_{1}\equiv\l$.
The term that remains when $\k_{0}=\k_{1}$ and that multiplies all higher
order terms in $\ve$ is an effective angular average over the dielectrics
$\e_{0},\e_{1}$.  This dominant term is {\it qualitatively different} from 
that arising from simple
surface tension renormalization $\g \rightarrow \g +{1\over 2}\s\vp =
\g+2\pi\s^{2}/(\e\k)$.  The higher order terms $S_{i}$ are determined with $\kappa_{1}$
relative to $\kappa_{0}$ and thus have their major effect when $\a \approx \pi$ as
more of the volume is occupied by electrolyte of inverse screening length $\kappa_{1}$.
For larger(smaller) $\l = \e_{0}/\e_{1}$, $S_{0}(S_{1})$ varies more significantly over
a smaller range of angles $\a$, reflecting the importance of the charged surface on the
lower dielectric slice. The effects of the liquid-liquid surface charge are symmetric
with the interchange $\a \rightarrow \pi-\a$, $\e_{0},\k_{0},\s_{0}\rightarrow
\e_{1},\k_{1},\s_{1}$ as expected.

{\it $\k_{0}/\k_{1}\rightarrow 0$ limit - }In the limit $\k_{1}\gg \k_{0}$
(\ref{KELVIN}) become $\Delta \vp_{0}= \vp_{0}$ and $\Delta\vp_{1}=\ve^{-2}\vp_{1}$,
where $\ve \equiv \kappa_{0}/\kappa_{1}$. Due to the strong screening in region 1, the
potential under the surface wet by liquid 1 vanishes as $\sim
\s_{1}(\e_{1}\kappa_{1})^{-1}$.  The first iteration makes the approximation
$\vp^{(0)}_{1}(r,\t)=f_{1}^{(0)}\approx 0$.  However, as a result of continuity of the
potential at $\theta=\a$, and finite $\kappa_{0}$ and $\s_{0}$, 

\begin{equation}
f_{0}^{(0)}(x,\t) = {2\pi^{2}\s_{0}\over \e_{0}\kappa_{0}x}
{\mbox{sh}x(\a-\t) \over \mbox{ch}x(\pi-\a)\mbox{ch}(\pi x/2)}.
\label{f00}
\end{equation}

\noindent Thus, $\vp_{0}(\theta>\a) = {\cal L}_{1}^{-1}f_{0}^{(0)} \neq 0$.
The first nonzero term in $\vp(P)$ arises from using (\ref{f00})
in the jump condition at $\theta=\a$:

\begin{equation}
\begin{array}{l}
\displaystyle \partial_{\theta}f_{1}(x,\a)-{\e_{0}\over \e_{1}}{\cal L}_{\ve}
{\cal L}^{-1}_{1}\partial_{\theta}f_{0}^{(0)}(x',\a)
=\ve\left({2\pi^{2}\s_{m}\over \e_{1}\kappa_{0}\mbox{ch}(\pi x/2)}\right).
\label{JUMPe1}
\end{array}
\end{equation}

\noindent Upon expanding $K_{ix}(\xi)$ about its dominant contribution 
at small $\xi$ in the operator 
${\cal L}^{-1}_{1}$ and performing the integration over $\xi$,

\begin{equation}
\begin{array}{l}
\displaystyle {\cal L}_{\ve}
{\cal L}^{-1}_{1}\partial_{\theta}f_{0}^{(0)}(x',\a)
= {\pi \s_{0}\over \e_{0}\kappa_{0}}\sum_{n=0}^{\infty}
{\ve^{2n}\over n!} \times \\[13pt]
\displaystyle\mbox{Im}\int_{0}^{\infty} {x'\Gamma\left(
n-{i\over 2}(x+x')\right)\Gamma\left(
n+{i\over 2}(x-x')\right)e^{-ix'\ln \ve} \over \mbox{ch}(\pi x'/2)
\mbox{ch} x'(\pi-\a) \Gamma(n+1-ix')}dx'
\label{INTEGRAL2}
\end{array}
\end{equation}

\begin{figure}
\includegraphics[width=3.2in]{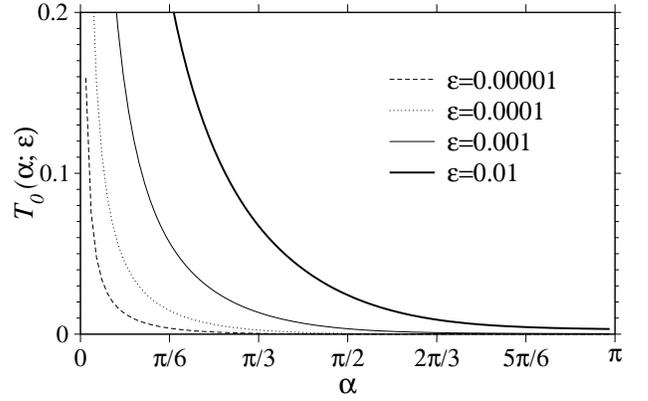}
\caption{The function $T_{0}(\a;\ve)$ (Eqn. (\ref{T})) that gives the 
$\s_{0}$-proportional term for the 
potential (Eqn. (\ref{VPT})) in the $\ve=\k_{0}/\k_{1}\rightarrow 0$ limit.}
\label{T}
\end{figure}

\noindent The contour integral (\ref{INTEGRAL2}) can be performed exactly to find the two lowest
order terms in $\ve$ (for $n=0$, $\mu^{(1)}(\ve) = \ve^{\pi/(2\pi-2\a)}$ and
$\mu^{(2)}(\ve) = \ve$) that must be considered. Using (\ref{VP0}), our final result is

\begin{equation}
\vp(P;\ve/\a\rightarrow 0)
\sim {2\pi^{2}\left(\s_{0}T_{0}(\a;\ve)+\s_{m}\ve
\right)\over\a \e_{1}\kappa_{0}} 
+ O\left(\ve^{{\pi\over \pi-\a}},\ve^{2}\right),
\label{VPT}
\end{equation}

\noindent where 

\begin{equation}
T_{0}(\a;\ve) \equiv {\pi \Gamma^{2}\left({\pi\over 4(\pi-\a)}\right)\ve^{\pi/(2\pi-2\a)}
\over 4(\pi-\a)^{2}\cos\left({\pi^{2}\over 4(\pi-\a)}\right)
\Gamma\left({3\pi-2\a\over 2\pi-2\a}\right)} - {\ve \over \cos\a}
\label{T0}
\end{equation}

\noindent is plotted in Fig. \ref{T} for various values of $\ve$.

The $\k_{0}/\k_{1}\rightarrow 0$ analysis fails for small $\a$, when the 
fully screened approximation breaks down as the interface approaches the 
$\s_{1}$-charged solid substrate. The correction to the potential is 
independent of $\s_{1}$ since it is screened out by a large $\k_{1}$. However, 
the jump condition across $\t=\a$ preserves an $\e_{1}$ dependence.  

We have explicitly incorporated the geometric dependence of long-ranged electrostatic
effects by deriving implicit equations for the contact angle valid for $R, L_{h}\sim
(\e/\s)\sqrt{\k\g_{01}} \gg \kappa^{-1}$.  Our results show that for large mismatch in
screening lengths, the potential at the contact point is proportional to the larger
screening length $\k_{0}^{-1}$ but varies as an $\a$-dependent power of the mismatch
$\k_{0}/\k_{1}$. In the $\kappa \approx \kappa_{1}$ case, the zeroth order
contribution to the potential can yield a nonnegligible effect:  For $\kappa \approx
50$nm, $\e_{0}=20, \e_{1}=80$, $\s_{0}-\s_{1} \approx 0.1$, $\s_{0},\s_{1} \sim O(1)$,
the zeroth order correction to the surface energy varies from $\sim 1.5 - 0.4
k_{B}T/\ell_{B}^{2}$ as $\a$ varies from $0$ to $\pi$.  Since one $k_{B}T/\ell_{B}^{2}
\simeq 8$ dyne/cm, the implicit $\a$ dependence should be observable as long as the
contact angle can be measured at a distance within 50nm of $P$. Such measurements are
possible using AFM techniques \cite{POMPE}. Geometry-dependent electrostatics may also
be an origin for the discrepancy between standard theory and measurements performed at
low salt concentrations when screening length are long \cite{FOKKINK}. 
Even when the
apparent contact angle cannot be measured within a screening length of $P$, our
results may provide a basis by which to better understand dynamic phenomena. Since
contact line pinning can occur at a length scale smaller than $\vert P-P^{*}\vert$,
the dynamics of the contact line may be better correlated with the true contact angle
$\a$ rather than $\a^{*}$.  

Note that $S_{i}(\a;\l)$ and $T_{0}(\a;\ve)$, when used in (\ref{RESULT1}) allow for
the possibility of {\it two} solutions for $\a$. The physical value for $\a$ will be
determined by the minimum energy solution that needs to be determined by the full
solution of the shape along $P-P^{*}$. Therefore, in cases where two roots for $\a$
are possible, we expect the minimum energy solution to be the one closest to $\a^{*}$.
Our results also imply that dramatic effects on wettability ($\a \approx 0,\pi$) can
be induced by small changes in the physical parameters.  

In our strictly 2D analyses, the only bounded solution when $\k_{0}=\s_{0}=0$ ({\it e.g.} air)  is
$\vp(P;\a)=0$. However, an asymptotic analysis for the large radius ($R \gg L_{h},
\kappa^{-1}$) limit is possible.  The correction term to the Y-D equation resulting from
such an analysis is proportional to $\tau \propto \s_{1}^{2}/(\k_{1}^{2}R)$ times an
$\a$-dependent factor, and defines the {\it line} tension \cite{POMPE,WHITE,LAW99} arising
from electrostatics.  These electrostatic forces giving rise to line tension may be
sufficiently long-ranged to be experimentally determined through a measurement of $\a$. 
Ionic strength may thus provide a controllable parameter with which to measure the elusive,
and controversial, line tension of wetting \cite{LAW99}.
 
The author thanks G. Bal and J. B. Keller for discussions, and 
the NSF for support via grant  DMS-9804370.

\end{document}